\documentclass[11pt]{article}
\usepackage{CJKutf8}
\usepackage{amsmath,amssymb,color,graphics,epsfig,tikz}

\DeclareMathOperator{\arctanh}{arctanh}
\usepackage{amsfonts}

\newcommand{\be}{\begin{equation}}
\newcommand{\ee}{\end{equation}}
\newcommand{\bea}{\setlength\arraycolsep{2pt} \begin{eqnarray}}
\newcommand{\eea}{\end{eqnarray}}
\newcommand{\nn}{\nonumber}

\def\fft#1#2{{\frac{#1}{#2}}}

\def\0{{\sst{(0)}}}
\def\1{{\sst{(1)}}}
\def\2{{\sst{(2)}}}
\def\3{{\sst{(3)}}}
\def\4{{\sst{(4)}}}
\def\5{{\sst{(5)}}}
\def\6{{\sst{(6)}}}
\def\7{{\sst{(7)}}}
\def\8{{\sst{(8)}}}
\def\sst#1{{\scriptscriptstyle #1}}

\begin{document}
\begin{CJK}{UTF8}{gbsn}

\begin{flushright}
\end{flushright}

\vspace{25pt}
\begin{center}
{\large {\bf Generalizing the exact quantization rule to multiple real turning points }}

\vspace{10pt}
Wei Yang

\vspace{10pt}

{\it College of Physics and Electronic Information Engineering, Guilin University of
Technology, Guilin, 541004, China}

\vspace{40pt}

\underline{ABSTRACT}
\end{center}
We generalize the exact quantization rule to multiple turning points, which are all on the real axis and are even in number.
We found that when we take wave functions of different energy levels, they are stable between two adjacent turning points and are always integer multiples of $\pi$.
We verify the effectiveness of the exact quantization rule by examining double square well potential and biharmonic potential.	
 
\vfill {\footnotesize Emails: weiyang@glut.edu.cn}
\thispagestyle{empty}

\pagebreak



\newpage

\section{Introduction}
\label{Sec.introduce}
The exact solutions to the Schr\"{o}dinger equation play crucial roles in quantum physics. It is well-known that there are only several potentials can be exactly solved, while most of them require some approximation methods to obtain the solutions.
It is well-known that there are several potentials can be exactly solved, for examples the
harmonic oscillator, the Coulomb, the Morse \cite{Morse}, P\"{o}schll-Teller \cite{Teller}, Eckart \cite{Eckart:1930zza} potentials, and these potentials can all be solved using exact quantization rules\cite{Ma:2005sy,Ma:2005}. 
This formula describes well how to obtain the correction term of WKB quantization when the potential energy has two turning points. It is expressed as follows
\begin{align}
\int_{x_A}^{x_B} k(x)dx=N\pi +\int_{x_A}^{x_B}\fft{k'(x)}{\log(\phi(x))'}dx \,. 
\label{eq:1}
\end{align}
Where $k(x)$ represents momentum for Schr\"{o}dinger equation, $x_1, x_2$ are two turning points, i.e. the points are zero of momentum.
Here $\phi(x)$ is the logarithmic derivative of the wave function $\psi(x)$.
$N$ is the number of nodes of $\phi(x)$.
The exact quantization rule is an exact expression of the Sommerfeld quantization condition and WKB  approximation methods\cite{Wentzel:1926aor,Kramers:1926njj,Brillouin:1926blg}, which is very helpful for us to deeply understand the bound state.
As the correction term generally does not depend on the nodes of the wave function, the equation  (\ref{eq:1}) can be simplified to proper quantization rule as follows \cite{19}.
\begin{align}
\int_{x_A}^{x_B} k(x)dx=n\pi -\int_{x_{0A}}^{x_{0B}}k_0(x)dx\,.
\label{eq:2}
\end{align}
where $n=N-1$ is the number of nodes in the wave function. 
The exact quantization rule method is a powerful tool
in finding the eigenvalues of solvable quantum systems \cite{20,21,22,23,24,25,26}.

In this paper, we generalize the exact quantization rule (\ref{eq:1}) to the case where all turning points are on the real axis.
We found that there is an similar equation in each potential well, and the turning points of different potential wells are connected by a quantum tunneling effect. 

This work is organized as follows. In Section \ref{Sec.2} we present the basic method in the double-well potential.
In Section \ref{Sec.3} we generalize to the Multi-well potential.
In Section \ref{Sec.4} we apply this method to Double square well potential.
In Section \ref{Sec.5} we apply this method to biharmonic potential.
In Section \ref{Sec.conclusion} we will give some conclusions for this paper.

\section{Double-well potential}
\label{Sec.2}
Consider the one-dimensional Schr\"{o}dinger equation
\begin{align}
[\frac{d^2}{d x^2} +E-V(x)]\psi(x)=0\,.
\label{eq:g2}
\end{align}
Here we ignore the coefficient $\hbar^2/2m$ for simplicity. 
The logarithmic derivative of the wave function is defined as $\phi(x)=\log(\psi(x))^{\prime}$, where the prime represent derivation for $x$, then the Schr\"{o}dinger equation become Riccati equation
\begin{align}
-\phi'(x)=k(x)^2+\phi(x)^2\,.
\label{eq:p1}
\end{align}
where $k(x)=\sqrt{E-V(x)}$, or
\begin{align}
-\phi'(x)=-\kappa(x)^2+\phi(x)^2\,.
\label{eq:p2}
\end{align}
where $\kappa(x)=\sqrt{V(x)-E}$.

In this case we consider a double-well potential, which is not necessarily symmetrical. 
Choosing the energy $E$ as a parameter and have 4 turning points with $V(x)$ on the real axis. We explicitly write out the potential $V(x)$ to satisfy
\begin{align}
V(x)\equiv
\begin{cases} 
&V(x)>E\qquad  -\infty<x<x_a\,, \\ 
&V(x)<E\qquad  x_a<x<x_b\,, \\
&V(x)>E\qquad  x_b<x<x_c\,, \\  
&V(x)<E\qquad  x_c<x<x_d\,, \\ 
&V(x)>E\qquad  x_d\leq x<\infty\,.
\end{cases} 
\end{align}  
In $-\infty<x\leq x_a$, the physically acceptable exponential decay solution
\begin{align}
\psi(x)\sim e^{\phi(x) x}\,,\qquad \phi(x)=\sqrt{V(x)-E}>0\,.
\label{eq:a}
\end{align}
In $x_d<x\leq \infty$, the physically admissible solution is
\begin{align}
\psi(x)\sim e^{\phi(x) x}\,,\qquad \phi(x)=-\sqrt{V(x)-E}>0\,.
\label{eq:d}
\end{align}
We find that the wave functions in both domains are exponentially decaying solutions, where no nodes exist.

In the region $x_a<x<x_b$, where $E \geq V (x)$.
Ma-Xu replaced the continuous potential well with a stack of thin films, where the potential energy of each layer is constant \cite{Ma:2005sy}. They derived the identity equation
\begin{align}
\int_{x_a}^{x_b}[k(x)-\fft{k(x)'}{(\log\phi)'}]dx
&=N_{ab}\pi-\arctan\fft{k_a}{\phi(x_a)}+\arctan\fft{k_b}{\phi(x_b)}\,.
\label{eq:ab}
\end{align} 
where $k_a=\sqrt{E-V(x_a)}, k_b=\sqrt{E-V(x_b)}$, 
the simplification uses equation (\ref{eq:p1}), and $N_{ab}$ is the number of zeroes of $\phi(x)$ in the region $x_a<x<x_b$.

Similar calculations can be performed in the region $x_c < x <x_d$, where $E \geq V (x)$. The identity equation write as
\begin{align}
\int_{x_c}^{x_d}[k(x)-\fft{k(x)'}{(\log\phi)'}]dx
&=N_{cd}\pi-\arctan\fft{k_c}{\phi(x_c)}+\arctan\fft{k_d}{\phi(x_d)}\,.
\label{eq:cd}
\end{align}  
where $k_c=\sqrt{E-V(x_c)}, k_d=\sqrt{E-V(x_d)}$, and $N_{cd}$ is the number of zeroes of $\phi(x)$ in the region $x_c<x<x_d$.

For region in $x_b<x<x_c$,  where $E \leq V (x)$.   
Using equation (\ref{eq:p2}), we only need to let $k(x)=i\kappa(x)$ in equation (\ref{eq:ab}) and modify the upper and lower limits of the integral, and write as
\begin{align}
i\int_{x_b}^{x_c}[\kappa(x)-\fft{\kappa(x)'}{(\log\phi)'}]dx
&=-N_{bc}\pi-i\arctanh\fft{\kappa_b}{\phi(x_b)}+i\arctanh\fft{\kappa_c}{\phi(x_c)}\,.
\label{eq:bc}
\end{align} 
where $k_b=\sqrt{V(x_b)-E}, k_c=\sqrt{V(x_c)-E}$, and $N_{bc}$ is the number of $\kappa(x) = \phi(x)$ in the region $x_b<x<x_c$.  
From equation (\ref{eq:p2}) we  can see
From the equation (\ref{eq:p2}), we can see that when $\kappa(x) > \phi(x)$, $\phi(x)$ increases monotonically with $x$, and when $\kappa(x) < \phi(x)$, $\phi(x)$ decreases monotonically with $x$. So $\phi(x)$ oscillates along $\kappa(x)$ in this region. When $E$ increases, $\kappa(x)$ gradually decreases, which means $\phi(x)$ oscillates along $\kappa(x)$ oscillates more frequently along k in this region.
The same method can be used by Ma-Xu to cut the continuous potential with a stack of thin films, using the concepts of mathematical analysis to obtain the identity equation (\ref{eq:bc}), See Appendix (\ref{Sec.appendix1}).
 
When $x$ is in the region $-\infty<x<x_a $ and $x_d < x <\infty$, the wave functions corresponding to these two regions are decaying, therefore wave function have no node in the region, so we can ignore this region when considering bound states.
 
If the potential is continuous, at $x=x_a,x=x_b,x=x_c,x=x_d$, we have $V(x)=E$, that mean $k_a=k_b=k_c=k_d=\kappa_b=\kappa_c=0$,  we have 
\begin{align}
&-\arctan\fft{k_a}{\phi(x_a)}+\arctan\fft{k_b}{\phi(x_b)}-i\arctanh\fft{\kappa_b}{\phi(x_b)}+i\arctanh\fft{\kappa_c}{\phi(x_c)}\nn\\
&-\arctan\fft{k_c}{\phi(x_c)}+\arctan\fft{k_d}{\phi(x_d)}=0 \,.
\label{eq:abc}
\end{align} 
So we can get the exact quantization rule for double-well potential with 4 turning points on the real axis by summing both sides of equations (\ref{eq:ab}),(\ref{eq:bc}),and (\ref{eq:cd}).
\begin{align}
&\int_{x_a}^{x_b}[k(x)-\fft{k(x)'}{(\log\phi)'}]dx+i\int_{x_b}^{x_c}[\kappa(x)-\fft{\kappa(x)'}{(\log\phi)'}]dx+\int_{x_c}^{x_d}[k(x)-\fft{k(x)'}{(\log\phi)'}]dx\nn\\
&=N\pi\,.
\label{eq:abcd}
\end{align} 
Where $N=N_{ab}-N_{bc}+N_{cd}$ is the total number of nodes of $\phi(x)$ in the entire real axis, or the total number $N-1$ of nodes of $\psi(x)$ in the entire real axis plus one node at infinity. 
We can see that quantum tunneling will reduce the number of nodes in the potential wells, or in other words, bound state particle in different potential wells can be transformed into each other through instantons. 

If the potential is not continuous, at $x=x_a,x=x_b,x=x_c,x=x_d$, where $V(x)$ is multivalued. In this case, (\ref{eq:abc}) is not zero, and it must be added to the right side of equation (\ref{eq:abcd}).
\section{Multi-well potential}
\label{Sec.3}
In this case we consider a Multi-well potential. 
Choosing the energy $E$ as a parameter and have $2n$ turning points on real axis. Write out the potential $V(x)$  
\begin{align}
V(x)\equiv
\begin{cases}
&V(x)>E\qquad   -\infty<x\leq x_1\,, \\ 
&V(x)<E\qquad  x_1<x<x_2\,, \\
&V(x)>E\qquad  x_2<x<x_3\,, \\  
&\cdots \\  
&V(x)<E\qquad  x_{2n-1}<x<x_{2n}\,, \\ 
&V(x)>E\qquad  x_{2n}\leq x<\infty\,.
\end{cases} 
\end{align}  
If the potential well is continuous, similar to usd the double-well potential method, we can obtain the exact quantization rule
\begin{align}
&\sum_{m=1}^{n}\int_{x_{2m-1}}^{x_{2m}}[k(x)-\fft{k(x)'}{(\log\phi)'}]dx+i\sum_{i=1}^{n-1}\int_{x_{2m}}^{x_{2m+1}}[\kappa(x)-\fft{\kappa(x)'}{(\log\phi)'}]dx=N\pi\,.
\label{eq:abcde}
\end{align} 
Where $N=N_{12}-N_{23}+N_{34}-N_{45}+\cdots$ is the total number of nodes of $\phi(x)$ in the entire real axis.
If the potential well is not continuous,
we need to add (\ref{eq:abcf}) to the right side of equation (\ref{eq:abcde}).
\begin{align}
&-\arctan\fft{k_1}{\phi(x_1)}+\arctan\fft{k_2}{\phi(x_2)}-i\arctanh\fft{\kappa_2}{\phi(x_2)}+i\arctanh\fft{\kappa_3}{\phi(x_3)}\nn\\
&\cdots -i\arctanh\fft{\kappa_{2n-2}}{\phi(x_{2n-2})}+i\arctanh\fft{\kappa_{2n-1}}{\phi(x_{2n-1})} -\arctan\fft{k_{2n-1}}{\phi(x_{2n-1})}+\arctan\fft{k_{2n}}{\phi(x_{2n})} \,.
\label{eq:abcf}
\end{align} 
\section{Double square well potential}
\label{Sec.4}
In this case, we consider a double square well potential given by
\begin{align}
V(x)\equiv
\begin{cases}
&V(x)=V_I  \qquad   -\infty<x<x_a\,, \\ 
&V(x)=0 \qquad  x_a<x<x_b\,, \\
&V(x)=V_0 \qquad  x_b<x<x_c\,, \\  
&V(x)=0 \qquad  x_c<x<x_d\,, \\ 
&V(x)=V_F \qquad  x_d<x<\infty\,.
\end{cases} 
\end{align}  
Where $V_I,V_F,V_0$ are all greater than 0, the minimum of them is denoted as $V_{min}$. Consider $0<E<V_{min}$, let $k=\sqrt{E},\kappa=\sqrt{V_0-E}$, as $k, \kappa$ does not depend on $x$, so $k(x)'=\kappa(x)'=0$. Form equation (\ref{eq:a}) (\ref{eq:d}), we have $\phi(x_a)=\sqrt{V_I-E}$, $\phi(x_d)=-\sqrt{V_F-E}$. 

Solving the  Schr\"{o}dinger equation in region $x_a<x<x_b$, we have wave function ignores a normalized factor  
\begin{align}
\psi(x) = \sin(kx+\delta_1)
\end{align}  
In region $x_b<x<x_c$, we have  
\begin{align}
\psi(x) = \sinh(\kappa x+\delta_2)
\end{align}  
In region $x_c<x<x_d$, we have  
\begin{align}
\psi(x) = \sin(k x+\delta_3)
\end{align}  
The matching conditions at the turning points $x_a,x_b,x_c$ and $x_d$, give
\begin{align} 
\begin{cases}
&1/\phi(x_a)=\tan(k_a x_a+\delta_1)/k_a\,, \\ 
&1/\phi(x_b)=\tan(k_b x_b+\delta_1)/k_b=\tanh(\kappa_b x_b+\delta_2)/\kappa_b\,, \\
&1/\phi(x_c)=\tanh(\kappa_c x_c+\delta_2)/\kappa_c=\tan(k_c x_c+\delta_3)/k_c\,, \\
&1/\phi(x_d)=\tan(k_d x_d+\delta_3)/k_d\,. 
\end{cases} 
\label{eq:pp}
\end{align}  
Where $k_a=k_b=k_c=k_d=k$, $\kappa_b=\kappa_c=\kappa$. Using the periodic equation $\tan(x+n\pi)=\tan(x),\tanh(x+i n\pi)=\tanh(x)$, and eliminating the parameter $\delta_1,\delta_2,\delta_3$ to obtain the quantization condition is just the exact quantization rule  
\begin{align}
&\int_{x_a}^{x_b}k(x)dx+i\int_{x_b}^{x_c}\kappa(x)dx+\int_{x_c}^{x_d}k(x)dx\nn\\
&=N\pi-\arctan\fft{k_a}{\phi(x_a)}+\arctan\fft{k_b}{\phi(x_b)}-i\arctanh\fft{\kappa_b}{\phi(x_b)}+i\arctanh\fft{\kappa_c}{\phi(x_c)}\nn\\
&-\arctan\fft{k_c}{\phi(x_c)}+\arctan\fft{k_d}{\phi(x_d)}\,. 
\label{eq:pp1}
\end{align} 
The bound state solution for this double square well potential can be found in quantum mechanics book.
Let $x_a= -2, x_b = -1, x_c = 1, x_d = 2$, and $V_I=100, V_F=101, V_0=100$. Form equation (\ref{eq:pp}), we have the energy levels $E_0=6.83296, E_1=26.9768, E_2=58.974, E_3=96.5517$.
Substituting the ground state energy into the equation (\ref{eq:pp1}), we can calculate $N_{ab}=1,N_{bc}=1,N_{cd}=1$, that total $N=1$ is corresponds to the ground state. Similarly, for the first excited state, we have $N_{ab}=2,N_{bc}=2,N_{cd}=2$ is corresponds to the first excited state. The highly excited state has similar properties.
\section{Biharmonic potential}
\label{Sec.5}
In this case, we consider the potential energy consisting of a biharmonic potential, which is given by
\begin{align} 
V(x)=
\begin{cases}
(x+\alpha)^2\,,\qquad -\infty<x \leq 0 \\ 
(x-\beta)^2-\gamma\,,\qquad 0<x<\infty\,. 
\end{cases} 
\end{align}   
Consider $0<E<\alpha^2$, and $\alpha>0$, where  $\alpha^2=\beta^2-\gamma$, there have four turning points $x_a=-\alpha-\sqrt{E},x_b=-\alpha+\sqrt{E},x_c=\beta-\sqrt{E+\gamma},x_d=\beta+\sqrt{E+\gamma}$. 
The  Schr\"{o}dinger equation in region $-\infty<x \leq 0$ is give
\begin{align}
[\frac{d^2}{d x^2} +E-(x+\alpha)^2]\psi_L(x)=0\,.
\end{align}
The physically admissible solution is 
\begin{align}
\psi_L(x)=D[(E-1)/2,\sqrt{2}(x+\alpha)]\,.
\end{align}
Where $D[a,b]$ is Parabolic cylinder function.
The  Schr\"{o}dinger equation in region $0<x<\infty$ is give
\begin{align}
[\frac{d^2}{d x^2} +E-(x-\beta)^2+\gamma]\psi_R(x)=0\,.
\end{align}
And the solution is 
\begin{align}
\psi_R(x)=D[(E-1+\gamma)/2,\sqrt{2}(x-\beta)]\,.
\end{align}
This two solutions must matching in $x=0$
\begin{align}
\log[\psi_L(x)]'|_{x=0}=\log[\psi_R(x)]'|_{x=0}\,.
\end{align}
Simplified to
\begin{align}
\alpha+\beta-\fft{H[(E+1)/2, \alpha]}{H[(E-1)/2, \alpha]}+\fft{H[(E+1+\gamma)/2, -\beta]}{H[(E-1+\gamma)/2, -\beta]}=0\,.
\end{align}
Where $H[a,b]$ is Hermite polynomials.
Let $\alpha=2,\beta=3,\gamma=5$, we can numerically solve the ground state energy and first few excited state energy levels  is given $E_0=-3.99311$, $E_1=-1.93114$, $E_2=0.286635$, $E_3=2.73417$, $E_4=5.44053$, $E_5=8.50148$.
 
In this case, When $0<E<4$ we can verify the exact quantization rule  
\begin{align}
&\int_{x_a}^{x_b}[k(x)-\fft{k(x)'}{(\log\phi)'}]dx+i\int_{x_b}^{x_c}[\kappa(x)-\fft{\kappa(x)'}{(\log\phi)'}]dx+\int_{x_c}^{x_d}[k(x)-\fft{k(x)'}{(\log\phi)'}]dx\nn\\
&=N\pi\,.
\label{eq:pp3}
\end{align} 
Substituting the $E_2$ into the equation (\ref{eq:pp3}), we can calculate $N_{ab}=1,N_{bc}=-1,N_{cd}=3$, that total $N=3$ is corresponds to the second excited state. Similarly, for the $E_3$, we have $N_{ab}=2,N_{bc}=-2,N_{cd}=4$ is corresponds to the third excited state.
Our numerical calculations also show that when we take wave functions of different energy levels, their equations (\ref{eq:pp3}) between two adjacent turning points are stable and are always integer multiples of $\pi$, which means that we can always take the wave function of a certain energy level as a reference in the calculation.
If all turning points of Schr\"{o}dinger equation are all on the real axis at the ground state energy level, the formula can be similarly simplified to proper quantization rule.
When $0>E$ or $E>4$ the complex turning points will occur, which may lead to a modification of the integral calculation produced by infinite subdivision, because at this time we may have multiple options for the upper and lower limits of the integral.
\section{Conclusion}
\label{Sec.conclusion}
In this work, we generalize the exact quantization rule to multiple turning points using the matching condition of logarithmic derivatives, These turning points are all on the real axis and have an even number. 
This exact quantization rule can obtain the bound state energy levels of Schr\"{o}dinger  equation. 
The validity of the quantization rule is verified by examining some examples that can exactly solve the Schr\"{o}dinger equation.
This method provides us with a new perspective to understand the Schr\"{o}dinger equation from the Riccati equation.
We speculate that for an odd number of turning points on the real axis, the exact quantization rule may also hold, which should correspond to resonance states. 
This method also has some shortcomings. When there are complex turning points, the equation (\ref{eq:abcde}) needs to be modified, which makes it necessary for us to expand the integral variable to the complex plane. This will be our next step of research.
\section*{Acknowledgement}
This work was supported by the Scientific Research Foundation of Guilin University of Technology under grant No. GUTQDJJ2019206.
 
\appendix
\section{Short range potential}
\label{Sec.appendix1}
Consider the region $x_b<x<x_c$ and divide it into $n$ equal films with width $d_n$,in the $j$th film, $x_b + jd_n-d_n \leq  x \leq  x_b + jd_n$, $V (x)$ is replaced with a constant potential $V_j$
\begin{align}
V_j=V(x_b+x_b + jd_n-d_n/2)\,,\qquad \kappa_j=\sqrt{V_j-E}\,. 
\end{align}
On two ends of the film, the logarithm derivatives $\phi_j$ and $\phi_{j-1}$ , which should match with the logarithm derivatives at the ends of the neighboring films,  
\begin{align}
\phi_{j}=\kappa_j\fft{\kappa_j\tanh(\kappa_j d_n)+\phi_{j-1}}{\kappa_j+\phi_{j-1}\tanh(\kappa_j d_n)}
\label{eq:g0}
\end{align}
If let $\kappa_j/\phi_{j-1}=\tanh(x)$, $\kappa_j/\phi_{j}=\tanh(y)$, then the equation (\ref{eq:g0}) become 
\begin{align}
\tanh(x+k_jd_n)=\tanh(y) 
\end{align}
As $\tanh(z+i m\pi)=\tanh(z)$, so we have
\begin{align}
\kappa_i d_n=-\arctanh(\fft{\kappa_i}{\phi_{i-1}})+\arctanh(\fft{\kappa_i}{\phi_{i}})+im\pi\,.
\end{align} 
Where $m$ we have
\begin{align}
m=
\begin{cases}
0\,, \qquad  \kappa_i \neq \phi_{i} \,\, \text{occurs} \, \text{in} \,\,x_b + jd_n-d_n \leq  x \leq  x_b + jd_n\\ 
1\,, \qquad  \kappa_i = \phi_{i} \,\, \text{occurs} \, \text{in} \,\, x_b + jd_n-d_n \leq  x \leq  x_b + jd_n\\ 
\end{cases} 
\label{eq:g1}
\end{align} 
Summing the equation (\ref{eq:g1}) from $i = 1$
to $n$, we obtain
\begin{align}
\sum_{i=1}^{n}\kappa_id_i &= iN\pi-\arctanh(\fft{\kappa_1}{\phi_{0}})+\arctanh(\fft{\kappa_1}{\phi_{1}})-\arctanh(\fft{\kappa_2}{\phi_{1}})+\arctanh(\fft{\kappa_2}{\phi_{2}})\nn\\
&-\cdots-\arctanh(\fft{\kappa_n}{\phi_{n-1}})+\arctanh(\fft{\kappa_n}{\phi_{n}})\,.
\label{eq:g2}
\end{align} 
Where $\phi_{0}=\phi(x_b)$, $\phi_{n}=\phi(x_c)$, and $N$ is the number of $\kappa(x) = \phi(x)$ in the region $x_b<x<x_c$. When $n$ goes to infinity, the sum equation (\ref{eq:g2}) will becomes an integral 
\begin{align}
\int_{x_b}^{x_c}\kappa(x)dx&=iN\pi-\arctanh(\fft{\kappa_1}{\phi(x_b)})+\arctan(\fft{\kappa_n}{\phi(x_c)})+\int_{x_b}^{x_c}\fft{\phi(x)\kappa(x)'}{\kappa^2-\phi^2}dx\,.
\end{align} 
Used equation (\ref{eq:p2}) we  can simplify
\begin{align}
i\int_{x_b}^{x_c}[\kappa(x)-\fft{\kappa(x)'}{(\log\phi)'}]dx
&=- N \pi-i\arctanh\fft{\kappa_b}{\phi(x_b)}+i\arctanh\fft{\kappa_c}{\phi(x_c)}\,. 
\end{align}

\end{CJK}

\begin{thebibliography}{99}

\bibitem{Morse}
Morse, Philip M.. “Diatomic Molecules According to the Wave Mechanics. II. Vibrational Levels.” Physical Review \textbf{34}, 57-64 (1929).

\bibitem{Teller}
 G. P\"{o}schll and E. Teller. Z. Physik, \textbf{83}, 143-151 (1933).

\bibitem{Eckart:1930zza}
C.~Eckart,
``The Penetration of a Potential Barrier by Electrons,''
Phys. Rev. \textbf{35}, 1303-1309 (1930)

\bibitem{Ma:2005sy}
Z.~Q.~Ma and B.~W.~Xu,
``Exact quantization rules for bound states of the Schroedinger equation,''
Int. J. Mod. Phys. E \textbf{14}, 599-610 (2005)

\bibitem{Ma:2005} 
Z.Q. Ma and B. W. Xu 2005 {\it Europhys. Lett.} {\bf 69} 685-691

\bibitem{Wentzel:1926aor}
G.~Wentzel,
Z. Phys. \textbf{38}, no.6, 518-529 (1926) 

\bibitem{Kramers:1926njj}
H.~A.~Kramers,
Z. Phys. \textbf{39}, no.10, 828-840 (1926)
 
\bibitem{Brillouin:1926blg}
L.~Brillouin,
Compt. Rend. Hebd. Seances Acad. Sci. \textbf{183}, no.1, 24-26 (1926)
 
\bibitem{19} Wen-Chao Qiang, Shi-Hai Dong 2010 {\it  EPL} {\bf 89} 10003

\bibitem{20} Dong, Shi Hai , M. Cruz-Irisson, 
journal of mathematical chemistry 50.4(2012):881-892.

\bibitem{21} 
Serrano, F. A. , X. Y. Gu , and S. H. Dong . 
Journal of Mathematical Physics 51.8(2010):082103-082103-16.

\bibitem{22} 
Qiang W. C. and Dong S. H., Phys. Lett. A, 363 (2007) 169.

\bibitem{23} 
Qiang W. C., Zhou R. S. and Gao Y., J. Phys. A:
Math. Theor., 40 (2007) 1677.

\bibitem{24} 
Dong S. H. and Gonzalez-Cisneros A., Ann. Phys.
(N.Y.), 323 (2008) 1136.

\bibitem{25} 
 Serrano, Fernando Adan , and S. H. Dong . 
 International Journal of Quantum Chemistry (2012).

\bibitem{26} 
Grandati, Y. , and A. Bérard. 
Physics Letters A 375.3(2011):390-395.

\end{thebibliography}
\end{document}